\documentclass[twocolumn]{webofc}
\usepackage[varg]{txfonts}   % Web of Conferences font
\usepackage{hyperref}
\usepackage{url}
\hypersetup{colorlinks=true,citecolor=blue,urlcolor=blue,linkcolor=blue}
\newcommand{\nuc}[2]{$^{{\mathrm{#1}}}${#2}}

\begin{document}
\title{Nuclear Reaction Data for Fission Products Off Stability}

\author{\firstname{Gustavo} \lastname{Nobre}\inst{1}\fnsep\thanks{\email{gnobre@bnl.gov}} \and
        \firstname{Emanuel} \lastname{Chimanski}\inst{1} \and
        \firstname{Aman} \lastname{Sharma}\inst{2}  \and
        \firstname{Alexander} \lastname{Voinov}\inst{3} \and
        \firstname{Kyle} \lastname{Wendt}\inst{2} \and
        \firstname{David} \lastname{Brown}\inst{1} \and
        \firstname{Shusen} \lastname{Liu}\inst{2}
        % etc.
}

\institute{Brookhaven National Laboratory 
\and
           Lawrence Livermore National Laboratory 
\and
           Ohio University
          }

\abstract{Neutron cross sections on fission products are relevant to a wide range of applications, including nuclear nonproliferation and forensics, spent-fuel assay, reactor burnup and design, as well as astrophysics. Evaluated nuclear data libraries generally fulfill application needs for isotopes on or near stability, however, for unstable fission products, theoretical descriptions of neutron-induced reactions often constitute the only available source of information. These models often make use of simplified assumptions, leading to unquantified impacts on predicted cross sections. In this work, we discuss possible approaches to addressing these issues, particularly by leveraging machine-learning methods, improved predictive reaction modeling, and experimental data to better constrain model parameters. 
Our goal is to eventually produce evaluated files for the most-produced nuclei off stability in the fission process of \nuc{235}{U} and submit them to the ENDF/B for consideration in the future ENDF/B-IX.0 release.
Here we present the methodology and discuss preliminary results comparing usual simplified approaches with a more realistic one accounting for nuclear deformation.

}
\maketitle

\section{Introduction}
\label{sec:Intro}

Accurate modeling of isotope generation and depletion are crucial for applications such as post-detonation forensics, nuclear energy (spent-fuel assay, reactor burnup and isotopics), radiation transport, and astrophysics. These require a complete description of neutron cross sections for fission products. On or near stability this is fulfilled by the libraries such as ENDF/B-VIII.1~\cite{VIII.1-full}, which is benchmarked against differential and integral experiments. Off stability, experimental data are either too scarce or non-existent, providing little guidance for evaluators. Therefore, reliable reaction model extrapolations must be used. Current alternatives present important limitations deeming them unusable for these purposes.

To address this, the present project, Realistic Reaction Evaluations for Fission Products Off Stability (RREFPOS), aims to use state-of-the-art reaction models, leveraging machine-learning advancements, and guidance level-density experiments to develop and disseminate complete neutron evaluations of fission-products off-stability, validating measured data for nuclei of interest for nonproliferation applications. 

\begin{figure}
\centering
	\includegraphics[scale=0.50, clip, trim= 0mm 0mm 0mm 0mm]{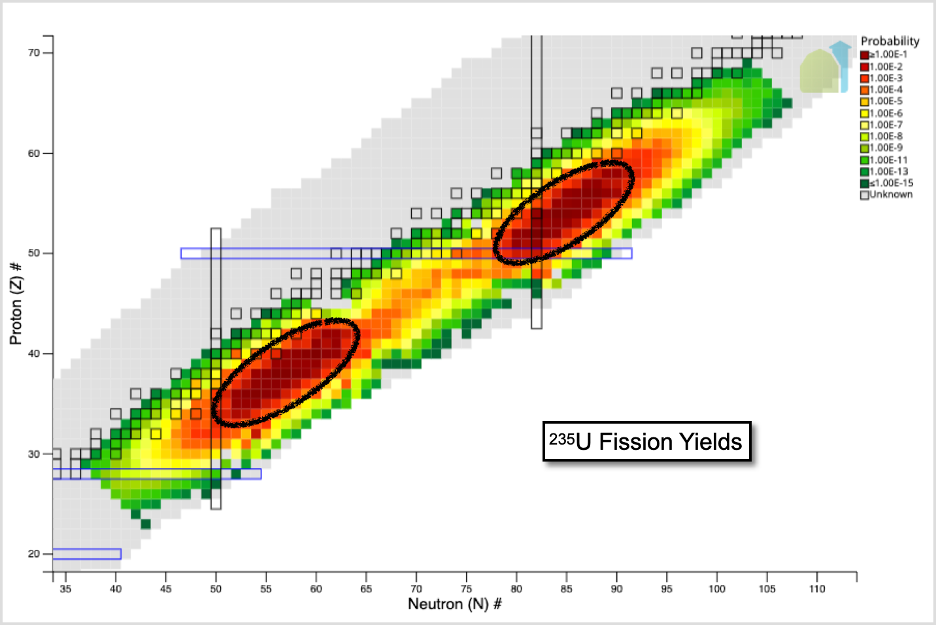}	
	\caption{Chart of nuclides indicating the nuclei relevant to the project primary goal, with isotopes color-coded by relative probability of formation through \nuc{235}{U} spontaneous fission.}
	\label{fig:nudat_235u_sfy}
\end{figure}

Improved iterations of the evaluations will be submitted to the ENDF/B library, allowing them to be potentially incorporated into future ENDF/B releases. This will also allow for ample testing by the nuclear data community, leading to feedback for improvements and fine-tuning of the evaluations.  
The primary focus will be in the nuclei mostly produced in \nuc{235}{U} spontaneous fission\footnote{Primary goal of RREFPOS -- 1$^\mathrm{st}$ Fission yield hump: \nuc{87-89}{Br}, \nuc{88-92}{Kr}, \nuc{91-94}{Rb}, \nuc{92-97}{Sr}, \nuc{95-99}{Y}, \nuc{97-102}{Zr}, \nuc{101-103}{Nb}; 2$^\mathrm{nd}$ Fission yield hump: \nuc{131-133}{Sb}, \nuc{132-136}{Te}, \nuc{135-138}{I}, \nuc{136-141}{Xe}, \nuc{139-143}{Cs}, \nuc{141-146}{Ba}, \nuc{144-145}{La}, \nuc{147-148}{Ce}}, shown in dark red within the black circle in Fig.~\ref{fig:nudat_235u_sfy}.

Once the methods are well established and the evaluation system is in place, a relatively small effort would then be required to extend the evaluation process to other unstable fission products. If that stage is reached early enough, we define as a secondary project goal to also produce evaluations for the mostly-produced unstable isotopes in the \nuc{239}{Pu} and \nuc{252}{Cf} spontaneous fission\footnote{Secondary goal of RREFPOS -- 1$^\mathrm{st}$ Fission yield hump: \nuc{94,100}{Y}, \nuc{96,103}{Zr}, \nuc{99,100,104,105}{Nb}, \nuc{102-108}{Mo}, \nuc{105-110}{Tc}, \nuc{107-112}{Ru}, \nuc{110-114}{Rh}, \nuc{112-116}{Pd}, \nuc{114}{Ag}; 2$^\mathrm{nd}$ Fission yield hump: \nuc{131}{Te}, \nuc{134}{I}, \nuc{135}{Xe}, \nuc{137,138,144}{Cs}, \nuc{140}{Ba}, \nuc{143,146-148}{La}, \nuc{145,146,149,150}{Ce}, \nuc{149-152}{Pr}, \nuc{151-153}{Nd}}. 
If time allows, as a stretch goal, we will also provide realistic evaluated files for the whole\footnote{Excluding nuclei for which on-stability data-constrained evaluations already exist in ENDF/B-VIII.1.} isotopic chain of fission products from \nuc{235}{U}, \nuc{239}{Pu}, and \nuc{252}{Cf}\footnote{Stretch RREFPOS goal -- \nuc{66}{V}, \nuc{66-67}{Cr}, \nuc{66-71}{Mn}, \nuc{66-75}{Fe}, \nuc{66-77}{Co}, \nuc{66-80}{Ni}, \nuc{66-82}{Cu}, \nuc{66-85}{Zn}, \nuc{68-87}{Ga}, \nuc{70-90}{Ge}, \nuc{72-92}{As}, \nuc{75-95}{Se}, \nuc{77-98}{Br}, \nuc{79-101}{Kr}, \nuc{81,83-103}{Rb}, \nuc{83-106}{Sr}, \nuc{87-109}{Y}, \nuc{88-112}{Zr}, \nuc{91-114}{Nb}, \nuc{93-117}{Mo}, \nuc{97-119}{Tc}, \nuc{98-121,124}{Ru}, \nuc{101-125}{Rh}, \nuc{103-126,128}{Pd}, \nuc{106-132}{Ag}, \nuc{108-134}{Cd}, \nuc{111-137}{In}, \nuc{113-139}{Sn}, \nuc{118-140}{Sb}, \nuc{120-143}{Te}, \nuc{123,125,126,128-145}{I}, \nuc{125,128,130-148}{Xe}, \nuc{131-151}{Cs}, \nuc{132-153}{Ba}, \nuc{135,137-155}{La}, \nuc{137-157}{Ce}, \nuc{139-159}{Pr}, \nuc{142-161}{Nd}, \nuc{144-163}{Pm}, \nuc{147-165}{Sm}, \nuc{149,151-168}{Eu}, \nuc{152-170}{Gd}, \nuc{155-172}{Tb}, \nuc{157-172}{Dy}, \nuc{161-172}{Ho}, \nuc{162-172}{Er}, \nuc{165-172}{Tm}, \nuc{168-172}{Yb}, \nuc{171-172}{Lu}}.

Evaluated files will be released in both legacy ENDF-6 and modern GNDS-2.0 formats and will satisfy requirements of main processing codes (NJOY~\cite{NJOY}, FUDGE~\cite{FUDGE2023}, AMPX~\cite{AMPX}).

\section{Planned Methodology}

The modeling of neutron-induced cross sections is normally divided in three regimes, (i) thermal, (ii) resonance, and (iii) fast regions.   In the first energy range, the world’s data is used to determine the best thermal cross section (e.g., Ref.~\cite{ATLAS2018}). In region (ii), if there are measurements and the resonances can be experimentally resolved individually, then the cross sections are determined by R-matrix fits to resonance parameters and we note a recent development to better characterize resonances using machine learning techniques~\cite{BRR}. If individual resonances cannot be individually resolved, then the goal becomes to describe the cross section probability distribution, with its average often extrapolated down from the fast range. In the fast region, fluctuations overlap so much that measured cross section is smooth and described by statistical models such as Hauser-Feshbach, in additional to direct-reaction formalism. Model parameters are then fit to reproduced available experimental data. In the cases of nuclei off stability, the lack or scarce availability of direct cross section data for nuclei off stability poses a challenge to this process in all energy regions. That is why reliance on more microscopic predictive models and novel approaches are of utmost importance for realistic evaluations.

\subsection{Resonance region}
For stable nuclei, the traditional approach to producing resonance region evaluations is to directly measure cross sections and fit the data with R-matrix parameters.  For unstable nuclei, there is no experimental data to fit.  The TENDL~\cite{TENDL} library approaches this by stochastically generating a set of resonances using estimated values of average resonance parameters -- widths ($\Gamma_\gamma$, $\Gamma_\mathrm{n}$) and spacings ($D_\lambda$).  However, the difference in cross section value near and far from the resonance peak can differ by several orders of magnitude. Considering the position of resonance peaks cannot be predicted, such resonance sampling can be meaningless unless averaged. Additionally, the TENDL library does not include the uncertainty in the average resonance parameters.

Aiming to improve upon such limitations, we will leverage a machine learning approach. Instead of directly using stochastically-generated sets if resonances, we will use them to establish the cross section probability distribution $P(\sigma|E,T)$ as a function of neutron energy $E$ and target temperature $T$.   
We note that our approach is similar to the approach taken to deal with unresolved resonances in reactor calculations. In the unresolved resonance problem, $P(\sigma|E,T)$ is directly measured from resonance set realizations for each isotope.  Here our eventual goal is a model that can predict $P(\sigma|E,T)$ as a function of $\Gamma_\gamma$, $\Gamma_\mathrm{n}$ and $D$ without generating sets of resonances.

\subsection{Fast-range reaction calculations}

Whenever there are dosimetry files~\cite{IRDFF} and/or experimental data in EXFOR~\cite{EXFOR} available, they should be reproduced by the new evaluated files. If there are no data available, which is expected to be the situation in the majority of cases, we will maximize the reliability of the calculations by employing machine-learning and microscopic models.

\subsubsection{Reaction model calculations}

For stable nuclei, reaction modeling, in combination with parameter fitting to observed data, fills the gaps where experimental data was not measured while providing a complete multi-reaction channel self-consistent description of the neutron-nucleus interaction. In the case of unstable nuclei, due to the scarcity or lack of direct measurements, predictive models are the only path towards more realistic evaluations. Therefore, in our calculations we give preference to microscopic models whenever possible, as they should have an expected uncertainty behavior when extrapolated.

A key ingredient in nuclear reaction calculations is the optical model potential (OMP), as it defines the total cross sections and the various transmission coefficients leading to compound-nucleus formation and prequilibrium emissions. For neutrons scattered from stable spherical or nearly-spherical nuclei, there is multitude of local and global OMPs, such as the Koning-Delaroche parameterization~\cite{KD}.

\begin{figure}
\centering
	\includegraphics[scale=0.50, clip, trim= 0mm 0mm 0mm 0mm]{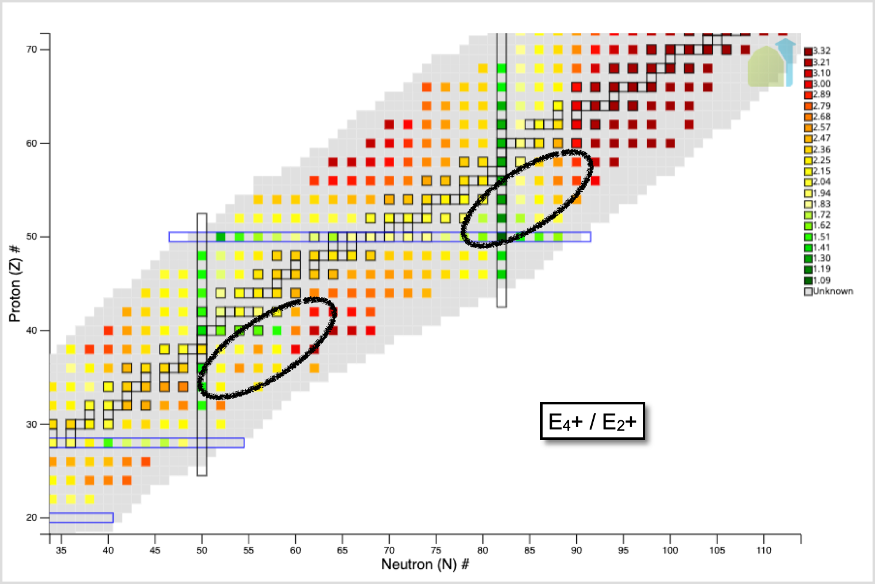}	
	\caption{Chart of nuclides indicating the nuclei relevant to the project primary goal, with isotopes color-coded by the value of the ratio of the excitation energy of the first 4$^{+}$ state by the first 2$^{+}$ one ($E_{4^+}$/$E_{2^+}$). This is an indication of deformation, as a value near 3.3 indicates a perfect rotor.}
	\label{fig:nudat_def}
\end{figure}

However, as seen in Fig.~\ref{fig:nudat_def}, most nuclei in the region of interest are quite well-deformed, rendering the spherical-nucleus OMP a severely poor approximation~\cite{NOBRE2014266}. To account for this, we will employ the the adiabatic model proposed in Refs.~\cite{PhysRevC.91.024618,NOBRE2014266,Nobre:2014,Herman:2014}, based on the work of Ref.~\cite{PhysRevC.85.044611}. This model has demonstrated to describe reasonably well statically-deformed nuclei, having as inputs simply the multipole deformation parameters ~\cite{PhysRevC.91.024618}. Refs.~\cite{PhysRevC.91.024618,NOBRE2014266}  show many examples of how this model can describe well the total cross section for well-measured statically-deformed nuclei and also how the elastic and inelastic angular distributions can be consistently well-reproduced, in contrast to to the simple interpolation of the spherical OMP.

Considering this approach has essentially  the multipolar deformation parameters and a regional potential as input, this model can more reliably be extended to nuclei off-stability by combining with nuclear deformations derived from microscopic models, such as the  HFB nuclear densities~\cite{Hilaire2007kb}.

\subsubsection{Machine-Learning for cross-section priors}

We will leverage the work done at Lawrence Livermore National Laboratory (LLNL) on developing a generative neural model of fast incident neutron-induced reaction cross section that is trained to learn systematic trends across the nuclear chart.  These trends are learned by training the network to translate the cross section of one nucleus to that of a neighboring nucleus on the nuclear chart, informed by density functional theory. 
By learning trends in nuclear data and conditioning those trends on extrapolated nuclear theory predictions, these networks will have a limited but non-trivial predictive power on nearby unstable isotopes.  This is possible because the network itself is never really extrapolating, but rather using predictions from density functional and related theories to adjust cross sections from nearby isotopes. Instead of using these outputs directly, they will then be used as informative priors to guide model parameters.    

\subsection{Level-density measurements}

The nuclear level density (NLD) models which are currently used in models within reaction codes such as EMPIRE~\cite{Herman:2007} are experimentally constrained only by the observed discrete levels and by the values of s/p/d-wave neutron resonance spacings (e.g., from Refs.~\cite{ATLAS2018,RIPL3}). 
This already corresponds to a very limited constraint to NLD on stable nuclei, as this information is limited to specific excitation energies, spins and parities. For nuclei off-stability, it becomes even more challenging as such data are seldom available.
As part of this project, we will  improve experimental constraints by performing targeted NLD measurements,  obtained through the particle evaporation technique suggested in Ref.~\cite{vonach_1983}  and currently developed at the Edwards Accelerator Lab~\cite{PhysRevC.99.054609}. This technique provides nuclear level density data in a wider excitation energy region (from the ground state up to the neutron separation energy) and broader range of spins. In this work we focus on the mass region related to fission product nuclei near the stability line, allowing for a more reliable parameter extrapolation. 

\section{Preliminary Results and Outlook}

On the NLD measurement front, we have completed two experimental campaigns. The first one involving 
\nuc{92+96}{Zr}(p,n) and \nuc{92}{Zr}(p,n) reactions led to measurements of the NLD of \nuc{92,96}{Nb}, while the second one, using the \nuc{94}{Zr}(p,n) reaction, helped to constrain the \nuc{94}{Nb} NLD. We have also performed neutron evaporation measurements of the \nuc{94}{Zr}(d,n) reaction, but its data analysis is still underway. This will eventually lead to data of \nuc{95}{Nb} NLD.

\begin{figure}
\centering
	\includegraphics[scale=0.335, clip, trim= 0mm 0mm 0mm 0mm]{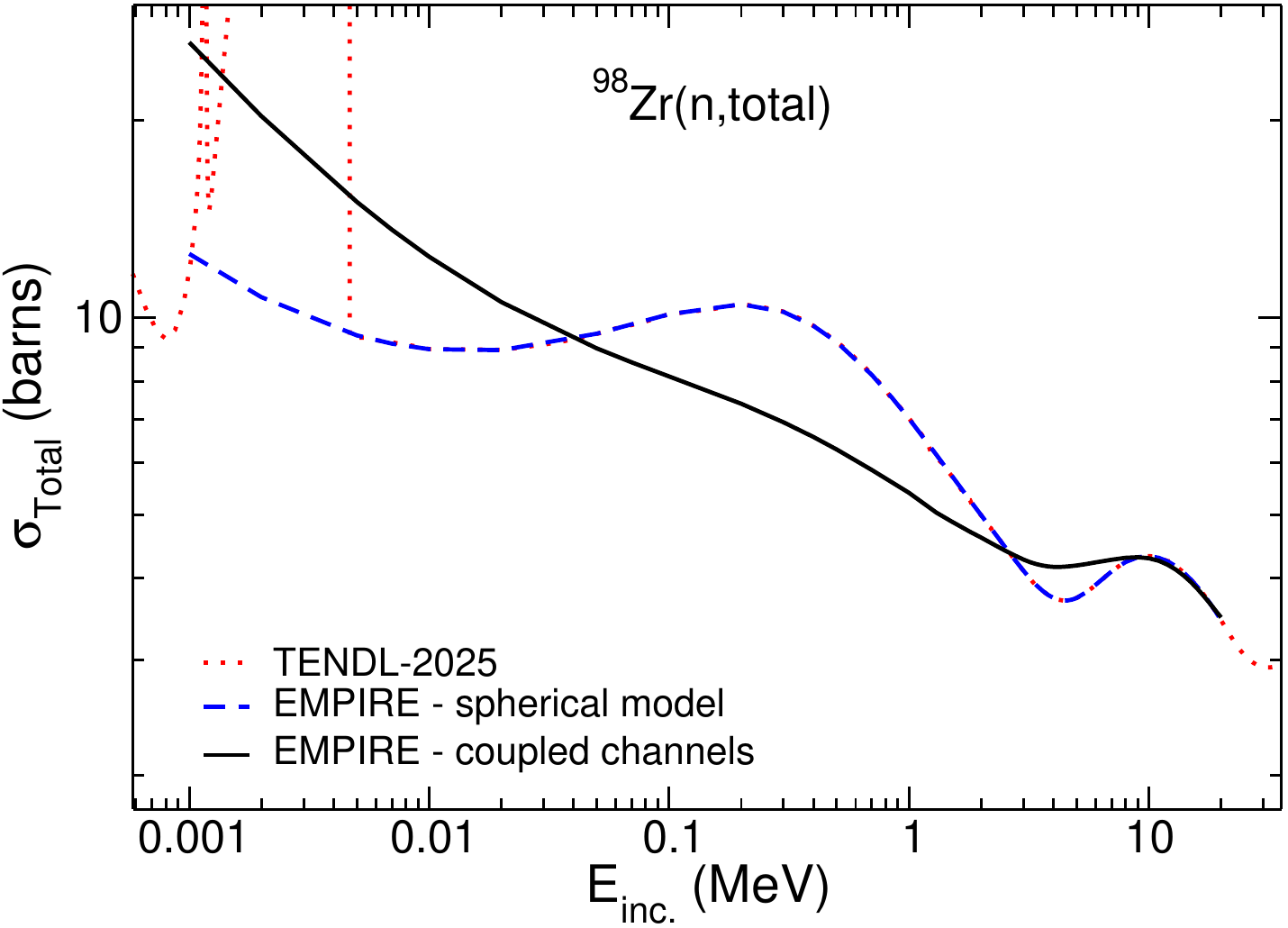}	
	\caption{Comparison of \nuc{98}{Zr}(n,total) cross section between a more realistic couple-channel calculation and a spherical one. TENDL also plotted as reference.}
	\label{fig:zr98-total}
\end{figure}

Regarding the reaction modeling, we have developed an infrastructure integrated to nuclear deformation databases,  allowing to run the EMPIRE~\cite{Herman:2007} reaction code within the adiabatic approach~\cite{PhysRevC.91.024618} for all relevant nuclei in the project primary goal. Preliminary calculations have been performed in the fast region and formatted into ENDF-6  files. As an example, Fig.~\ref{fig:zr98-total} shows preliminary results for the \nuc{98}{Zr}(n,total) cross section, comparing the  adiabatic coupled-channels approach, obtained  by deforming a global spherical OMP, with the direct use of the same spherical OMP. The two approaches differ substantially from one another. The cross section from TENDL, which closely follows the spherical model, is alsoshown . The \nuc{98}{Zr} is expected to have a strong dynamic quadrupole deformation, of the order of 0.33~\cite{RIPL3}. Therefore, incorporating deformation effects in the reaction treatment is expected to provide a significantly more realistic description than spherical approximations.

Going forward, we expect to further refine our fast-region calculations as well as focus on the developing our proposed approach for the resonance region. We plan to submit preliminary complete evaluated files to the ENDF/B library in the near future.

\section*{Acknowledgements}
This work was supported by the Office of Defense Nuclear Nonproliferation Research and Development within the U.S. Department of Energy’s National Nuclear Security Administration.
%
% BibTeX or Biber users please use (the style is already called in the class, ensure that the "woc.bst" style is in your local directory)
 \bibliography{Nobre-RREFPOS} % Replace "your_bib_file" with the actual name of your .bib file
%\begin{thebibliography}{}
%Journal Author, Article title. Journal \textbf{Volume}, page numbers (year). \url{https://doi.org/Article-DOI-number}
%Book Author, \textit{Book title} (Publisher, place, year) page numbers
% etc
%\end{thebibliography}
\vspace{-1cm}
\end{document}